\documentclass[reprint,nofootinbib,amsmath,amssymb,aps]{revtex4-1}
\usepackage{graphicx,dcolumn,bm,hyperref,float}
\usepackage{anyfontsize}
\usepackage{color}
\usepackage{xcolor}

\newcommand{\beq}{\begin{equation}}
\newcommand{\eeq}{\end{equation}}
\newcommand{\bea}{\begin{eqnarray}}
\newcommand{\eea}{\end{eqnarray}}
\newcommand{\bef}{\begin{figure}}
\newcommand{\eef}{\end{figure}}

\newcommand{\Egrav}{E_{\mbox{\tiny{gravity}}}}
\newcommand{\Eelec}{E_{\mbox{\tiny{electric}}}}
\newcommand{\Teff}{T_{\mbox{\tiny{eff}}}}
\newcommand{\Tbh}{T_{\mbox{\tiny{BH}}}}
\newcommand{\Tds}{T_{\mbox{\tiny{dS}}}}
\newcommand{\Nds}{N_{\mbox{\tiny{dS}}}}
\newcommand{\meff}{m_{\mbox{\tiny{eff}}}}

\begin{document}

\title{Inconsistency with De Sitter Spacetime  \\ in a New Approach to Gravitational Particle Production }

\author{Mark P.~Hertzberg$^{1}$}
\email{mark.hertzberg@tufts.edu}
\author{Abraham Loeb$^2$}
\email{aloeb@cfa.harvard.edu}
\affiliation{$^1$Institute of Cosmology, Department of Physics and Astronomy, Tufts University, Medford, MA 02155, USA
\looseness=-1}
\affiliation{$^2$Department of Astronomy, Harvard University, 60 Garden Street, Cambridge, MA 02138, USA
\looseness=-1}

\begin{abstract}
We study a claimed new mechanism for particle production and black hole evaporation through a spatially dependent temperature. This new temperature is comparable to the Hawking result near the black hole, but is very small far away, and therefore could be a small correction. Here we apply the proposed reasoning to the case of de Sitter space, finding that it over predicts the de Sitter temperature of a minimally coupled scalar by factor of $\approx 4.3$ and over predicts the particle production rate by a factor of $\approx 52$. For non-minimally coupled scalars, it has other various problems; it predicts a negative particle production for conformal, or nearly conformal, coupled scalars; it predicts unsuppressed productions of heavy scalars. This all demonstrates an inconsistency in the proposed formalism.
\end{abstract}

\maketitle


{\em Introduction and Review}.---
In a recent Physical Review Letter \cite{Wondrak:2023zdi} a new mechanism for massive objects to evaporate by emission of particles was proposed. In that work the authors make use of the imaginary part of the effective action. They provide an analogy to the Schwinger effect in electrodynamics whereby charged particles can be created in pairs in an external electric field and use this to argue that other particles can be created in pairs in an external gravitational field (not necessarily at a horizon). By replacing the square of the electromagnetic field strength by a combination of the square of the Riemann tensor, Ricci tensor, and Ricci scalar, they suggest one can just replace the Schwinger formula by a gravitational formula. 

In particular, they start with the characteristic energy for a particle produced via the Schwinger process in an electric field 
$\Eelec=[q^2 {\bf E}^2]^{1/4}$.
Then recognize that the factor of $q^2{\bf E}^2$ can be re-written (in the absence of magnetic fields) as 
\beq
\Eelec=[\Omega_{\mu\nu}\Omega^{\mu\nu}/2]^{1/4}
\eeq
where $\Omega_{\mu\nu}$ is the gauge-field curvature $\Omega_{\mu\nu}=iqF_{\mu\nu}$. By utilizing the effective action for scalar particles in curved space time, they note that the combination $\Omega_{\mu\nu}\Omega^{\mu\nu}/2$ appears in a way analogous to several contractions of Riemann, which they identify. Through this analogy, and using an effective action $\mbox{Im}[W]_{\rm eff}={1\over384\pi}\int d^4x\sqrt{-g}\, \Egrav^4$, they suggest that a gravitational field should also produce pairs of particles with characteristic energy
\beq
\Egrav=[(R_{\mu\nu\alpha\beta}R^{\mu\nu\alpha\beta}-R_{\mu\nu}R^{\mu\nu})/30+3(\xi-1/6)^2R^2]^{1/4}
\eeq
where $\xi$ is the non-minimal coupling parameter for a scalar (defined through $-\xi R\phi^2/2$ in the action); $\xi=0$ corresponds to minimally coupled and $\xi=1/6$ corresponds to conformally coupled. Ref.~\cite{Wondrak:2023zdi} focuses on the vacuum case (Schwarzschild black holes) in which $R_{\mu\nu}=0$ and $R=0$, which simplifies this expression; but we shall track the full expression here.

The next step is to connect to a temperature. Ref.~\cite{Wondrak:2023zdi} suggests the spectrum is thermal and identifies this characteristic energy with the average energy of a particle in the black body spectrum, namely $\langle E\rangle=(\pi^4/(30\zeta(3)))T$. Hence it is argued that the effective temperature of the scalars in a gravitational field is 
\beq
\Teff={30\zeta(3)\over\pi^4}\Egrav
\eeq

By applying this to the case of a Schwarzschild black hole (using $R_{\mu\nu\alpha\beta}R^{\mu\nu\alpha\beta}=48 G^2M^2/r^6$), the effective temperature is found to be radius, $r$, dependent, with $\Teff\sim \sqrt{GM}/r^{3/2}$. For $r$ an $\mathcal{O}(1)$ factor outside the horizon $2GM$, this effective temperature is $\Teff\sim 1/(GM)$, on the same order as the Hawking temperature $\Tbh=1/(8\pi G M)$ \cite{Hawking:1975vcx}. However, since it has a strong $r$ dependence as $\Teff\propto 1/r^{3/2}$ it is small at large distances. Hence in this example, the proposal is not changing the temperature measured at asymptotically large distances and so it is not obviously violating well established results. 

However, in this work we apply their methods to another system, namely de Sitter space, showing that here it clearly does violate established results. 

\begin{figure}[t]
    \centering
    \includegraphics[width=\linewidth]{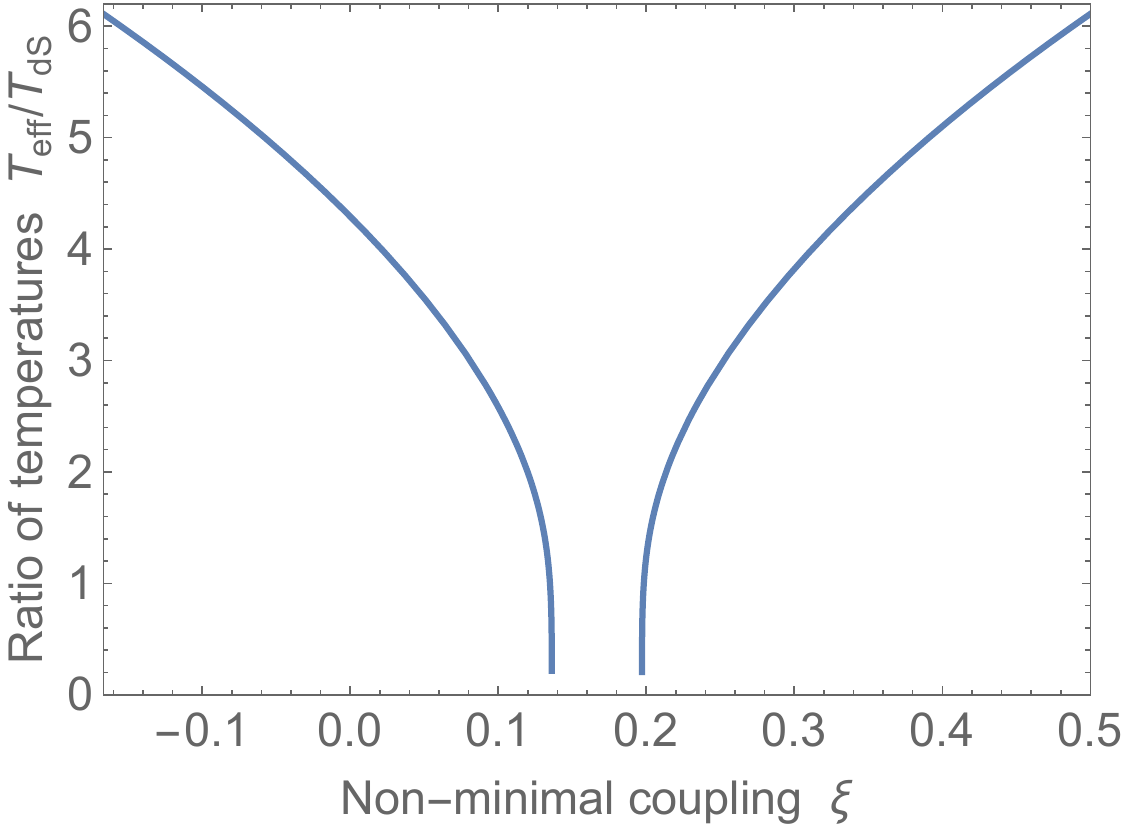}
    \caption{The ratio of temperatures $\Teff/\Tds$ as a function of non-minimal coupling $\xi$.}
    \label{FigRatio}
\end{figure}

{\em Application to de Sitter Space}.---
Since the results of Ref.~\cite{Wondrak:2023zdi}  are of a very general character, claiming that any gravitational field may potentially give rise to particle production, we wish to test their claims against other established results. Here we focus on de Sitter space; a maximally symmetric spacetime with $SO(4,1)$ symmetry. Denoting the (constant) de Sitter Hubble scale $H$, one can readily compute contractions of Riemann, to give $R_{\mu\nu\alpha\beta}R^{\mu\nu\alpha\beta}=24 H^4$, $R_{\mu\nu}R^{\mu\nu}=36 H^4$, and $R^2=144 H^4$. By inserting this into the above proposal for $\Egrav$ and then for $\Teff$, one obtains the effective temperature of the scalar in de Sitter space
\beq
\Teff={12\,2^{1/4}\,5^{3/4}\zeta(3)(-1+30(1-6\xi)^2)^{1/4}\over\pi^3}{H\over 2\pi}
\eeq
We can compare this to the Gibbons-Hawking \cite{Gibbons:1977mu} temperature of de Sitter space
\beq
\Tds={H\over 2\pi}
\eeq
We see that this new temperature $\Teff$ is parametrically of the same order, albeit with a different prefactor that depends on the non-minimal coupling $\xi$. The ratio of these two temperatures is plotted in Figure \ref{FigRatio}.

For the case of a minimally coupled scalar ($\xi=0$), we have
\beq
{\Teff\over\Tds} = {12\,5^{3/4}\,58^{1/4}\zeta(3)\over\pi^3}\approx 4.3
\eeq
Hence, this predicts that a scalar in de Sitter space would be considerably hotter than the standard Gibbons-Hawking result. Since the behavior of a minimally coupled scalar in de Sitter space has been studied carefully in numerous ways, for example one can just directly solve the mode function equations and obtain the power spectrum, it is well known that the value of $\Tds=H/(2\pi)$ is in fact correct. Importantly, there is no limit that can be taken for this new effect to become sub-dominant; this is unlike the black hole case where one could simply go to large distances to recover the Hawking result. Since de Sitter space is homogenous, such a limit is not possible; hence the discrepancy persists in all regimes. We believe this casts serious doubt on the result of Ref.~\cite{Wondrak:2023zdi}.

In the case of a conformally coupled scalar ($\xi=1/6$), the above effective temperature $\Teff$ is complex (this is in the part of Figure \ref{FigRatio} where there is no curve). The interpretation of this is unclear. One might claim that this means there is no particle production, which would be consistent with the conformal symmetry in this limit. Or it might just imply a further inconsistency. In any case, we find that for $\xi<0.135$ and $\xi>0.198$, the proposal has $\Teff$ real and greater than $\Tds$, which does not seem physical.

{\em Particle Production Rate}.---
We can also directly examine particle production. The prescription of Ref.~\cite{Wondrak:2023zdi} is to take the above $\Egrav$ and integrate it over a region to obtain the rate of produced particles as
\beq
{dN\over dt}=
{1\over192\pi}\int d^3x\sqrt{-g_3}\,\Egrav^4
\eeq
To apply this to de Sitter space, we integrate over the Hubble volume Vol $=4\pi/(3 H^3)$ and again use the values of $R_{\mu\nu\alpha\beta}$ for de Sitter to obtain
\beq
{dN\over dt}={H\over360}(29-360\xi+1080\xi^2)
\eeq
In Ref.~\cite{Wondrak:2023zdi} the focus was on the Schwarzschild spacetime, where there are $\mathcal{O}(1)$ correction factors due to the breaking of translation symmetry near the black hole. But since de Sitter space carries translation symmetry, there are no such correction factors. 

\begin{figure}[t]
    \centering
    \includegraphics[width=\linewidth]{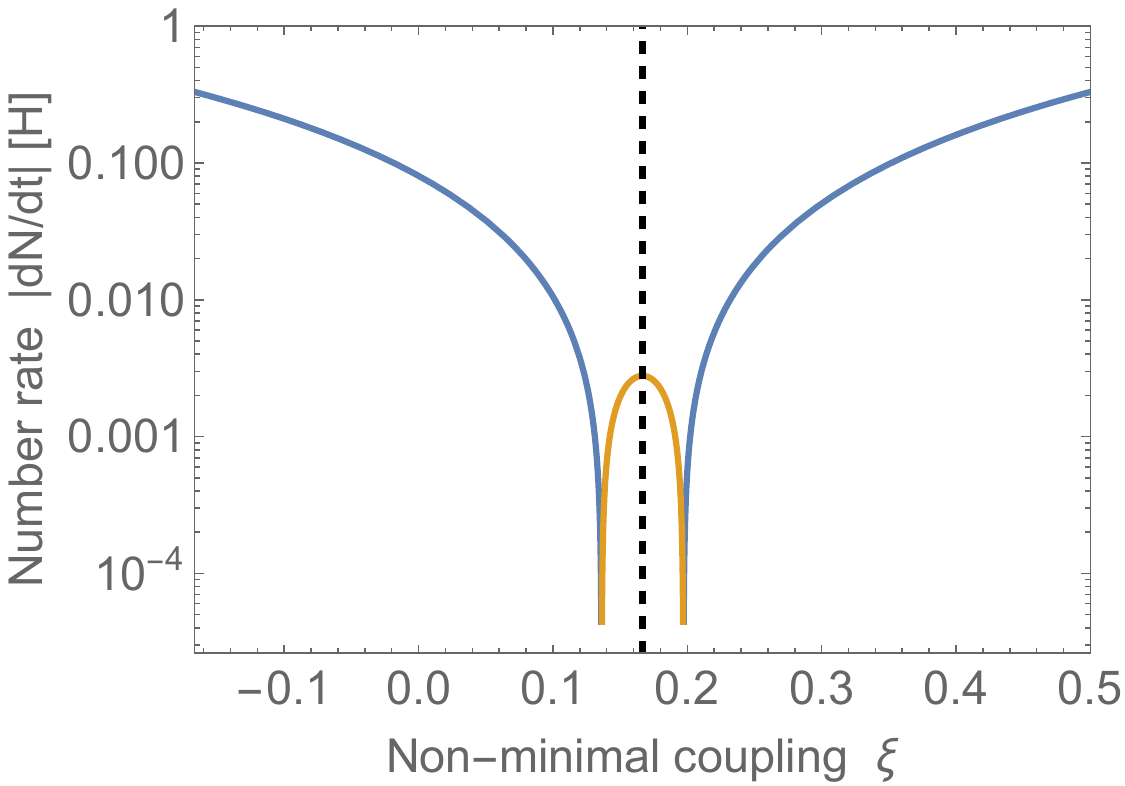}
    \caption{Magnitude of particle production rate $|dN/dt|$ (in units of $H$) as  a function of non-minimal coupling $\xi$. The blue curve indicates
    positive values of $dN/dt$, while the orange curve indicates negative values of $dN/dt$. }
    \label{FigNumber}
\end{figure}

We plot this rate in Figure \ref{FigNumber}. 
For the blue part of the curve $dN/dt>0$, while in the orange part of the curve $dN/dt<0$. 
In particular, for the case of conformal coupling $\xi=1/6$, this is negative. Hence the analysis of Ref.~\cite{Wondrak:2023zdi}, predicts that a conformally coupled scalar has a {\em negative} particle production. This looks to be very  unphysical. 

Let us now consider the case of a  minimally coupled scalar ($\xi=0$). Using the known result that de Sitter space has a 
temperature of $\Tds=H/(2\pi)$, one can compute the particle production rate as
\beq
{d\Nds\over dt}={A\over 4}\int {d^3p\over(2\pi)^3}{1\over e^{p/\Tds}-1}
\eeq
Where the factor of $1/4$ accounts for angular projection of emitted flux. Here $A$ is the area of the Hubble sphere  $A=4\pi/H^2$. Carrying out this integral gives
\beq
{d\Nds\over dt}={\zeta(3)\,H\over 8\pi^4}
\eeq
The ratio of these rates is
\beq
{dN\over dt}\Big{/}{d\Nds\over dt}={29\,\pi^4\over 45\,\zeta(3)}\approx 52
\eeq
Hence this new prescription over-predicts the particle production for a minimally coupled scalar in de Sitter space by a factor of $\approx 52$. 

For general values of $\xi$, it is well known that the scalar aquires an effective mass of
\beq
\meff^2=\xi\,R=12\, \xi\,H^2
\eeq
The mode functions $v_k$ for the scalar field are well known to obey the differential equation
\beq
v_k''+\left[k^2-\left(2-{\meff^2\over H^2}\right){1\over\eta^2}\right]v_k=0
\eeq
where $\eta=-e^{-Ht}/H$ is the conformal time variable. For the conformal value $\xi=1/6$, we have $\meff^2=2H^2$ and so this
reduces to $v_k''+k^2 v_k=0$, as in flat space, and there is precisely zero particle production. Furthermore, for large positive values of $\xi$, we have $\meff\gg H$ and the particle production (which can be expressed in terms of Hankel functions) becomes exponentially suppressed.

 Neither of these features is present in Figure \ref{FigNumber} for the claimed new approach to particle production in gravitational fields. 
Similar remarks apply to the energy production rate $dE/dt$. Altogether, this contradicts well established facts.

{\em Conclusions}.---
New approaches in gravitational evaporation are important to consider. However, the proposal of Ref.~\cite{Wondrak:2023zdi}, which indicates there can be particle production in any gravitational field, even away from black hole horizons, leads to new predictions for other spacetimes, including de Sitter spacetime. It predicts (i) an especially hot de Sitter space, (ii) a negative production for conformally coupled scalars, (iii) an unsuppressed production of heavy particles (from the non-minimal coupling). All of which is incompatible with established results.

{\em Note added}.---
While we wrote our paper, another work appeared in Ref.~\cite{Ferreiro:2023jfs} which also finds inconsistencies in the proposal. They examine the case of electromagnetism in more detail and comment on FLRW, while we examine de Sitter spacetime and the subtle dependence on $\xi$ in more detail (its complexity in conformal case, its precise value for minimally coupled, etc).

In the original version of Ref.~\cite{Ferreiro:2023jfs}, it is suggested that there is particle production in generic FLRW spacetimes for the conformally coupled scalar. However this claim has 2 shortcomings: (i) The proposal of Ref.~\cite{Wondrak:2023zdi} cannot obviously be directly used for time dependent spacetimes, which includes generic FLRW spacetimes. This is why in our work we have focussed on one of the few nontrivial cosmological spacetimes which is stationary, namely de Sitter. (ii) For the conformally coupled scalar, there is {\em not} in fact particle production, as we showed here. Instead, as we first showed, for $\xi=1/6$ one instead has $\Egrav^4$ (or $\Teff^4$) being {\em negative}, corresponding to a complex value when taking the fourth root. It is only for $\xi$ outside of a certain window (see Figure \ref{FigRatio}) that the fourth root is real and so there is in fact particle production.

Later, after our first version appeared, Ref.~\cite{Ferreiro:2023jfs} put out a new version which included an appendix. In this appendix, they realized these two shortcomings in their original version and corrected them: (i) they mentioned the special case of de Sitter space and (ii) the fact that the production is negative for the conformally coupled scalar. (To show this, they found $\mbox{Im}[W]_{\rm eff}=-{12\over 192\pi}\int d^4x\sqrt{-g}\,H^4$. While the minus sign is correct, which follows from the complex values for $\Egrav$ we originally discovered, the prefactor of $12/(192\pi)$ is incorrect; the correct value is $1/(960\pi)$, which can be  deduced from our above value for $\Egrav$ or $\Teff$.)

We summarize some of the central new points in our work: 
(i) the novelty of using de Sitter space (due to its stationarity), 
(ii) the precise computation for the action and the particle rates as a function of $\xi$, including the precise identification of positive versus negative production, or real versus complex effective energy, 
(iii) the identification that at large $\xi$ the approach would lead to a significant rate for an effectively heavy scalar in de Sitter space, rather than exponential suppression.

\bigskip

\section*{Acknowledgments}
We thank Georgios Fanaras for comments.
M.~P.~H.~ is supported in part by National Science Foundation grants PHY-2013953 and PHY-2310572. 
A.~L.~ is supported in part by the Black Hole Initiative at Harvard University which is funded by grants from the John Templeton Foundation and the Gordon and Betty Moore Foundation.


\end{document}